# Experimental Verification of Overlimiting Current by Surface Conduction and Electro-osmotic Flow in Microchannels

Sungmin Nam[1*‡], Inhee Cho[1*], Joonseong Heo[2], Geunbae Lim[2],
Martin Z. Bazant[3], Gunyong Sung[4†] and Sung Jae Kim[1,5†]

[1]*Department of Electrical and Computer Engineering,*
*Seoul National University, Seoul 151-744, Republic of Korea*
[2]*Department of Mechanical Engineering,*
*Pohang University of Science and Technology, Pohang 790-784, Republic of Korea*
[3]*Department of Chemical Engineering and Mathematics,*
*Massachusetts Institute of Technology, Cambridge 02139, USA*
[4]*Department of Material Science and Engineering,*
*Hallym University, Chunchon, 200-702, Republic of Korea*
[5]*Inter-university Semiconductor Research Center,*
*Seoul National University, Seoul 151-744, Republic of Korea*

[*]These authors contributed equally.
[†]Correspondence should be addressed to Prof. Sung Jae Kim and Prof. Gunyong Sung:
E-mail: (SJKim) gates@snu.ac.kr; phone: +82-2-880-1665 and (GSung) gysung@hallym.ac.kr; phone: +82-33-248-2361.
[‡]Current affiliation is Department of Mechanical Engineering, Stanford University, USA.

**ABSTRACT**

Possible mechanisms of overlimiting current in unsupported electrolytes, exceeding diffusion limitation, have been intensely studied for their fundamental significance and applications to desalination, separations, sensing, and energy storage. In bulk membrane systems, the primary physical mechanism is electro-convection, driven by electro-osmotic instability on the membrane surface. It has recently been predicted that confinement by charged surfaces in microchannels or porous media favors two new mechanisms, electro-osmotic flow (EOF) and surface conduction (SC), driven by large electric fields in the depleted region acting on the electric double layers on the sidewalls. Here, we provide the first direct evidence for the transition from SC to EOF above a critical channel height, using *in situ* particle tracking and current-voltage measurements in a micro/nanofluidic device. The dependence of the over-limiting conductance on channel depth (*d*) is consistent with theoretical predictions, scaling as $d^1$ for SC and $d^{4/5}$ for EOF with a transition around *d*=8um. This complete picture of surface-driven over-limiting current can guide engineering applications of ion concentration polarization phenomena in microfluidics and porous media.

**INTRODUCTION**

Over the past decade, electrokinetic phenomena in nanoscale fluidic channels have drawn significant attention, for both fundamental theory and novel engineering applications [1-6]. Much progress has been made in understanding electro-convection during ion concentration polarization (ICP) due to electro-osmotic flows near the membrane or nanochannel interface driving salt depletion [5, 7-10]. Due to the complexity of direct numerical simulation of the Poisson equation (for electric potential), Nernst-Planck equations (for ion concentrations) and Navier-Stokes equations (for fluid flows) in multi-dimensional geometries [9, 11, 12], as well as inherent limitations of the classical dilute solution model [13], it is crucial to directly observe in particle motion and flow fields in precisely controlled micro/nanofluidic geometries [14, 15].

Recent experimental investigations based on the micro/nanofluidic platform reveal complex electrokinetic phenomena in microchannels near an ion perm-selective membrane or nanochannel junctions, which cannot be described by one-dimensional diffusion-drift equations. The classical theory of ICP predicts a constant concentration gradient in the quasi-neutral electrolyte and saturation of the current to the Nernst diffusion-limited value at high voltage [16]. The key features of steady ICP under direct current are as follows. (1) In case of cation selective membrane, the electrolyte concentration at the anodic side of the membrane is strongly depleted within ICP layer and approaches zero at the membrane at the limiting current [17]. (2) Due to the low salt concentration, the electrical conductivity zone significantly decreases, leading to a greatly amplified local electric field [18]. (3) The large electric field drives fast electrokinetic flow inside the depletion zone leading to strong vortices in order to satisfy the continuity conditions [19, 20]. (4) The strongest vortex at the membrane leads to secondary vortices to form multiple concentration zones inside the depletion zone, providing inherent instability issues [21-

23]. (5) The limiting current and over-limiting conductance can be adjusted by manipulating the strong convection [8, 24]. Since the ion depletion zone expands with the strong convection, suppressing the convection can reduce the total electrical resistance of the system, although this trend depends on the voltage and geometry (since electro-convection clearly lowers the resistance close to the limiting current and compensates for the reduced conductivity). (6) The salt concentration during ICP in microchannels tends to form very sharp gradients between the depleted and concentrated regions, perhaps first observed a decade ago [25]. In micro/nano/micro-channel junctions, where steady over-limiting current has been observed [26], salt gradients propagate as shock waves, [6, 27] or ``deionization shocks'' [28-30] at constant current, due to the nonlinear effect of ion transport in the electric double layers of the sidewalls. These observations suggest that multiple transport mechanisms may be involved when over-limiting current occurs under strong confinement.

A unified theory of over-limiting conductance through an electrolyte confined within a charged microchannel has recently been developed [31, 32], but direct experimental confirmation is still lacking. The theory predicts a transition between two new mechanisms, surface conduction (SC) and electro-osmotic flow (EOF), that dominate in nanochannels and microchannels, respectively. The EOF mechanism is driven by large electro-osmotic slip in the depleted region on the sidewalls (not the membrane at the end of the channel) [33], leading to "wall fingers" of salt transported by vortices faster than transverse diffusion [31, 34]. This new mode surface convection thus cannot be described by classical Taylor-Aris dispersion [33, 34]. The EOF mechanism, extended for "eddy fingers" in a random porous medium, has been confirmed indirectly by experiments measuring the current-voltage relation, scalings with salt concentration and surface charge, and desalination efficiency of "shock electrodialysis" [35]. The SC mechanism has also been confirmed in straight nanopores with controlled surface charge by again predicting the current-voltage relation and by *ex situ* imaging of metal electrodeposits grown along the pore walls by surface conduction [36].

In this letter, we provide the first *in situ* observation of the SC and EOF mechanisms and the predicted geometrical transition between them. The motion of fluorescent tracer particles is visualized to reveal the internal dynamics in both regimes. The over-limiting conductance is also measured, and the predicted scalings with channel depth are confirmed, including a maximum that had escaped notice at the critical thickness of the transition.

Micro-nanofluidic devices are fabricated by surface patterning method [37] in PDMS substrate as shown in Figure 1. Current is driven in aqueous 1mM KCl solution through a cation permselective nafion nanojunction to generate ICP. Under the experimental conditions, the surface charge of the PDMS microchannel is negative, as expected, since EOF is observed to be flowing toward cathode. Each device has the same bulk microchannel conductance by fixing the cross-sectional area as the depth is varied, since the resistivity of bulk electrolyte is linearly proportional to the area. The physical dimensions are shown in Table 1. According to the theory [31], using typical surface charge in water, the dominant mechanism of overlimiting current mechanism should vary with the microchannel depth between surface conduction for $d<2$um, electro-osmotic surface convection for $2$um$<d<20$um, and electro-osmotic instability on the membrane for $d>20$um. The quasi-steady current-voltage relation is measured by linear sweep voltammetry with a slow sweep rate at 0.2V/15sec from 0V to 7V. A customized LabView program automatically records the current data at each step.

Below the limiting current, as expected, the current-voltage relations of the mirochannels with different depths all collapse onto a linear relationship, as shown in Figure 2, indicated by "range

(i)". The curves of 2um, 6.5um, 14.5um and 22um are only shown for the graphical simplicity. We refer to this as the "Ohmic region" since there is a constant apparent conductivity for steady electro-diffusion, even though both diffusion and electromigration of the cations contribute to the total flux [16]. As the salt concentration approaches zero at the membrane interface, the classical diffusion limited current is reached, as indicated by "range (ii)" in the Figure 2. As the applied voltage increases further, another region of over-limiting current with a smaller, constant conductance is observed (range (iii)), whose physical origin should be determined. The trend with depth is qualitatively similar to the theoretical predictions [31]. For the thinnest channel (2um), the limiting current range is almost missing, and the over-limiting conductance (slope) is larger than for all the larger depths. Moreover, the 2um depth leads to the largest over-limiting current, since it has the largest area to volume ratio, and thus the greatest effect of the new surface transport mechanisms.

In order to extract a quantitative conclusion, the overlimiting conductance for each depth is obtained by linear regression of the slope of the data in range (iii) and plotted in log-log scale as shown in Figure 3. The error bars increase in the deeper microchannels due to the flow instability, as clarified in the inset of Figure 3. While previous studies have established the simple power-law scalings of over-limiting conductance with salt concentration and surface charge in a fixed geometry [35, 36], our data reveal a non-monotonic dependence on the microchannel depth with a minimum conductance at a depth of roughly 8um. Although this non-trivial behavior follows from the theory [31, 35], it has not been recognized or studied until this work. In nanochannels, the over-limiting conductance due to surface conduction is proportional to the volume/area ratio, or inverse depth, $d^{-1}$, and thus decreases with increasing depth. In contrast, the over-limiting conductance due to electro-osmotic surface convection in microchannels is predicted to have the opposite trend, scaling as $d^{4/5}$. Although the 4/5 exponent follows from subtle scaling arguments [31, 35], the increasing conductance with increasing depth can be easily understood as a result of larger vortices carrying more convective flux, *i.e.* thicker wall fingers.

The theoretical scalings are consistent with the "V shaped" data in Figure 3. The decreasing $d^{-1}$ scaling for the smaller nanochannels is clear from the data and supports the SC theory. The increasing $d^{4/5}$ scaling for EOF is also consistent with the data for the larger microchannels as a possible limiting scaling law, although the data is not conclusive. Remarkably, however, the minimum conductance around $d$=8um is identical to the theoretical prediction of the critical depth from the intersection of the two scaling laws [31].

In order to correlate the scaling transition in overlimiting conductance with the existence of electro-convection, the electrokinetic flows are imaged during the current-voltage measurement, and their snapshots are shown in Figure 4. Each column has the same depth, and each row represents the time-evolution of the flow field. Note that the scale bar in the first column is different from others since 2um deep device has larger region of interest than others. The electrolyte contains a fluorescent dye and microparticles to track both concentration and flow fields, respectively. The length of depletion zone is also tracked by the dye, and it presumably represents the thickness of diffusive layer. (See the supporting information.).

As shown in the fourth row of the Figure 4, electrokinetic flows are significantly different and changed as a function of the depth of the microchannel, as clearly seen in the supporting videos for each depth. At 2um depth, a flat depletion zone propagates as a deionization shock wave from the nanojunction over the microchannel [6, 25, 27-30] and the strong vortical motions are largely suppressed and hardly observed because of geometrical constrictions [19, 26]. Transverse diffusion of the dye across the depth also eliminates concentration gradients [31, 34]. Closer to the transition depth (8um), a pair of point-like weak vortices are initiated at the top/bottom of microchannel in case of 6.5um deep microchannel, especially after the limiting current range (>2.5V). As the depth increases to 14.5um, the vortices are strengthen and become the primary convections at the side of microchannel, while they induce secondary vortices at the stagnation point in the middle of the microchannel, leading to a strong electro-convective flow [21]. Finally, the electro-convection becomes very strong and unstable in case of the 22um deep microchannel, thus approaching the bulk behavior [6, 13, 28-30]. Combining the results of the overlimiting conductance measurement and the flow field tracking (*e.g.* the transition from the surface conduction governed regime to the electro-convection governed regime), we can suggest that the mechanism of overlimiting current behavior cannot be attributed solely to surface conduction or electro-osmotic surface or bulk convection. Instead, each phenomenon plays an important role and dominates in different micro/nanochannel geometries.

In summary, we experimentally demonstrate the competition between different transport mechanisms for overlimiting current in microchannels by fixing (or changing) the conductance. With the microscopic imaging of electrokinetic flow and electrochemical measurements, we can separate the effects of the EOF and SC depending on the geometrical confinement. Consistent with the theory [31, 32], SC dominates in nanochannels and electro-osmotic surface convection in microchannels with a minimum conductance at the predicted transition around 8um depth. A clear understanding about the mechanism of the overlimiting current would be essential not only for scientific fundamentals but also furnishing effective engineering strategies to exploit and control ICP (and the overlimiting current) in electrochemical systems such as fuel cells, batteries, electro-desalination systems [2, 7, 8], and template-assisted electrodeposition [36].

**ACKNOWLEDGEMENTS**

This work is supported by Basic Science Research Program (2013R1A1A1008125), the Center for Integrated Smart Sensor funded as Global Frontier Project (CISS- 2011-0031870) and Future based Technology Development Program (Nano Fields) (2012-0001033) by the Ministry of Science, ICT & Future Planning and Korean Health Technology RND project, Ministry of Health and Welfare Republic of Korea (HI13C1468, HI14C0559).

**FIGURE CAPTIONS**

**Figure 1** Schematic diagrams of (a) H-shaped micro/nanofluidic device and (b) microscope image of the device. External electric voltages are applied at two north reservoirs, while two south reservoirs are electrically grounded. The depth ($d$) and height ($h$) are varying to obtain different cross-sectional area ($A=hd$) and unit boundary length ($L=2(h+d)$).

**Figure 2** Current-voltage relations for micro/nanofluidic devices with varying $h$ and $d$, but fixed cross-sectional area and thus fixed bulk microchannel conductance. As a result the conductance is constant for all devices below the limiting current, while limiting and overlimiting current values are significantly different. Voltage is swept at 0.2V/15sec. Each line is measured at least 10 times for guarantee a repeatability.

**Figure 3** Overlimiting conductance as a function of the depth of microchannel in log-log scale. Linear scaled plot is shown in inset. The data is consistent with the predicted transition from the surface conduction governed regime ($\propto d^{-1}$) to the electro-convection governed regime ($\propto d^{4/5}$).

**Figure 4** Microscopic images of flow tracking by fluorescent dye and particles. Each image is taken at a given time (or voltage) as noted. Note that the scale bars are different. Shallower depth effectively suppresses an electro-convective flow than deeper one. At $d=14.5$um, the convective flow at top and bottom creates a secondary vortical flow at the stagnation point in the middle of channel. Over $d=22$um, the flow finally becomes unstable. See supplementary videos.

# TABLE CAPTION

**Table 1** Physical dimensions of micro/nanofluidic device of the same cross-sectional area.

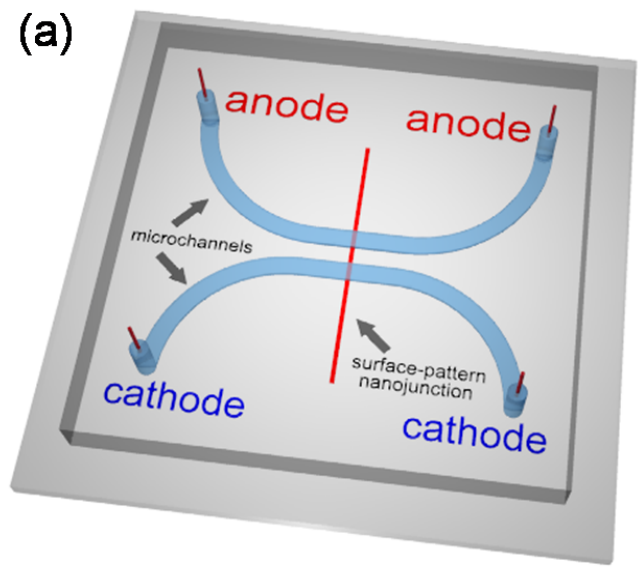

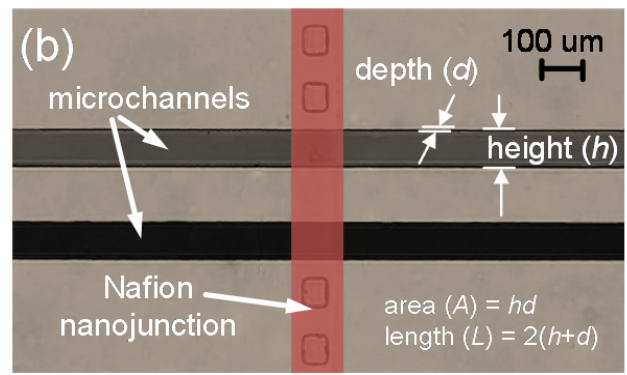

Figure 1

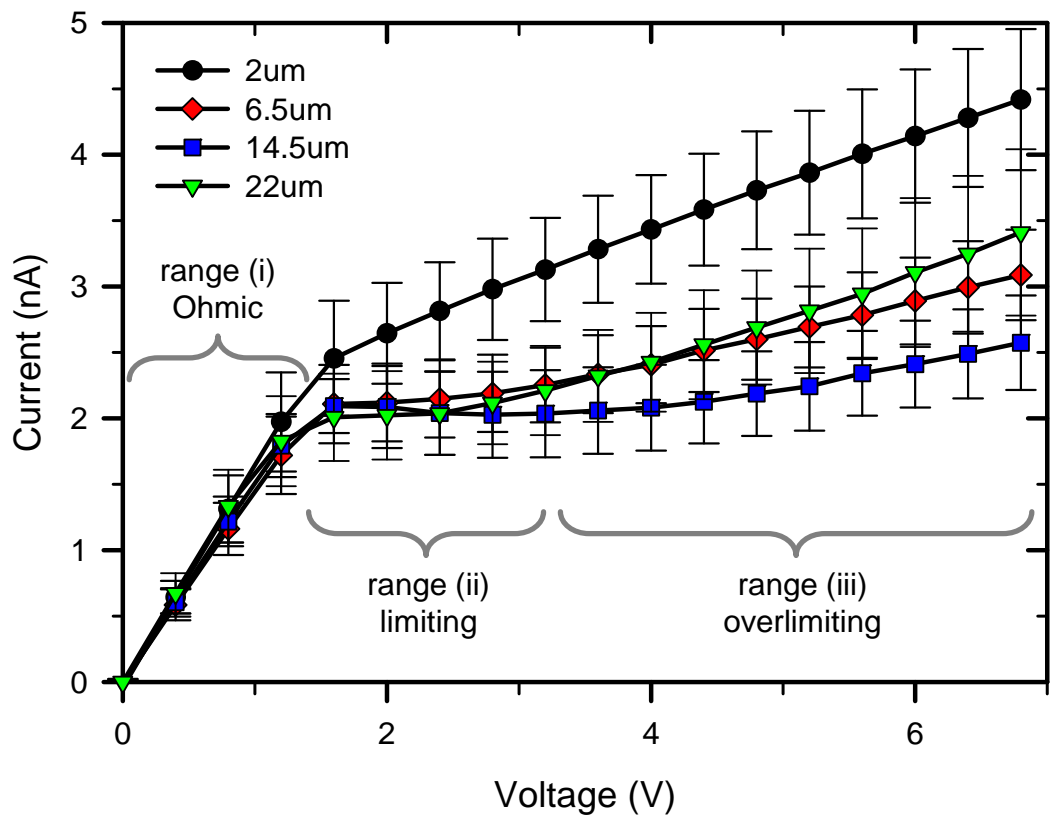

Figure 2

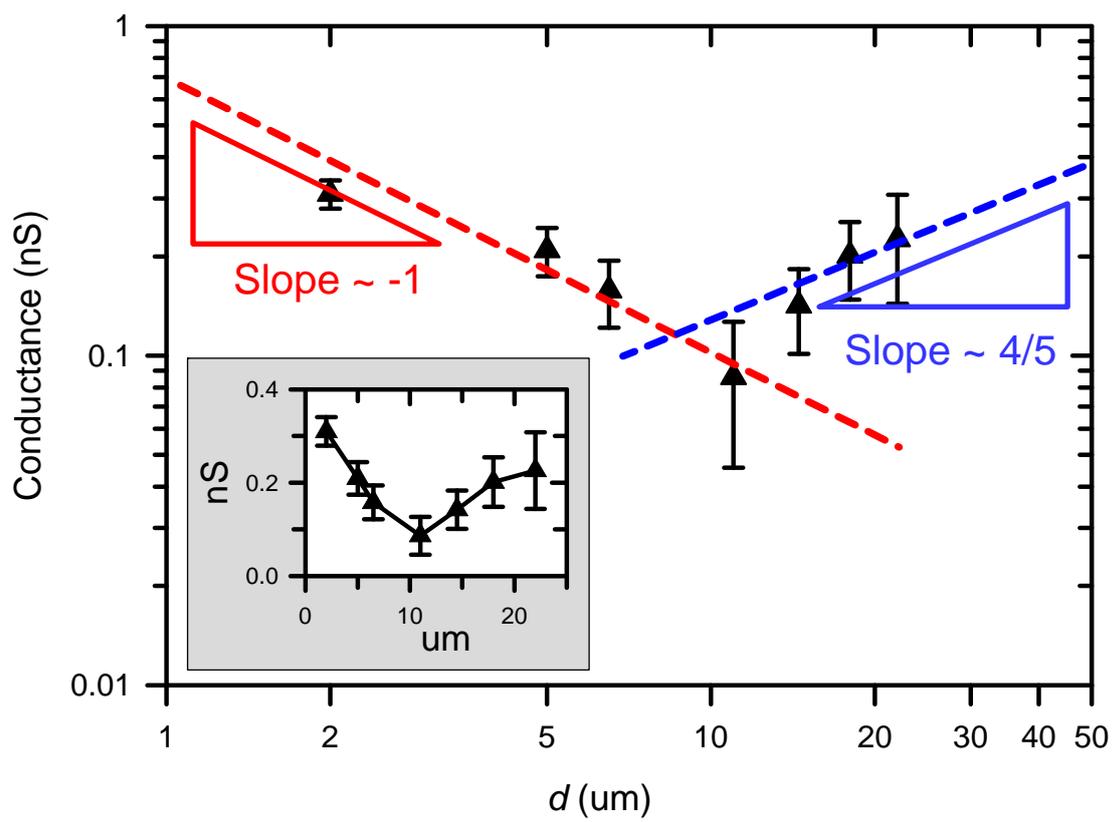

Figure 3

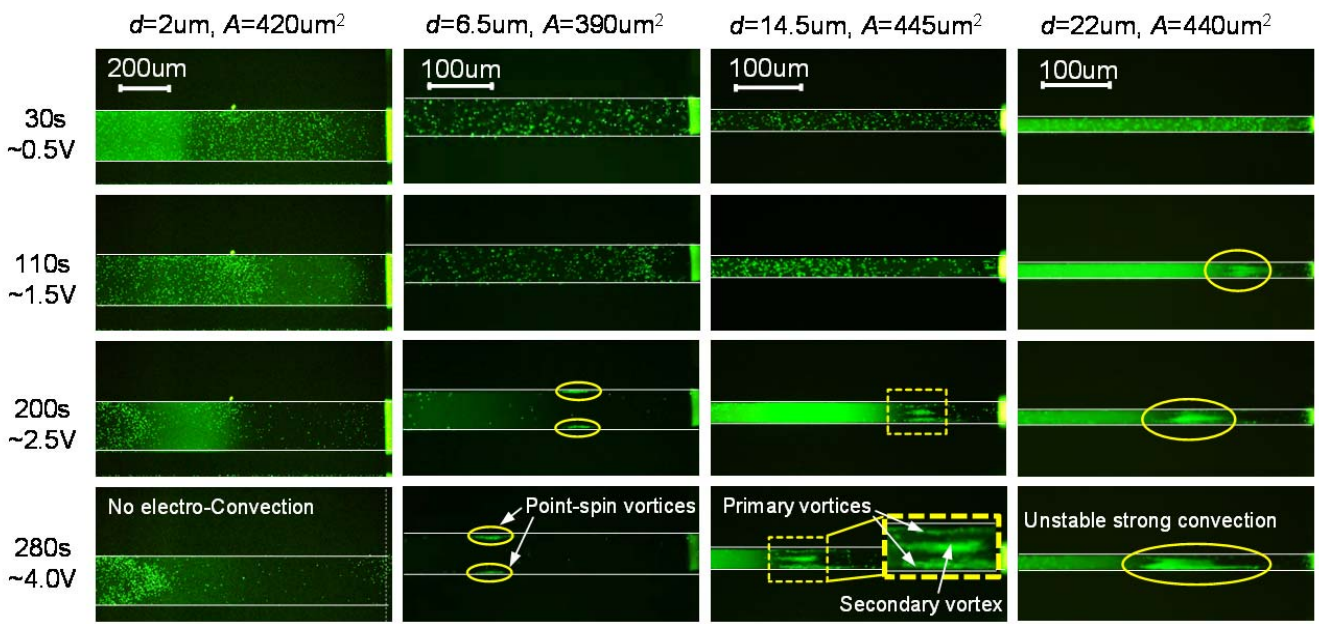

Figure 4

| *d* (um) | *h* (um) | *A* (um²) | *L* (um) |
|---|---|---|---|
| 2 | 210 | 420 | 424 |
| 5 | 84 | 420 | 178 |
| 6.5 | 60 | 390 | 133 |
| 11 | 38 | 418 | 98 |
| 14.5 | 30 | 435 | 89 |
| 18 | 23 | 414 | 82 |
| 22 | 20 | 440 | 84 |

Table 1